\documentstyle[psfig,aps]{revtex}
\def\m@thcombine#1#2{%
  \setbox0=\hbox{$#1$}
  \setbox1=\hbox{$#2$}
  \ifdim\wd0>\wd1
    \setbox0=\hbox to\wd1{\hss\box0\hss}
  \else
    \setbox1=\hbox to\wd0{\hss\box1\hss}
  \fi
  \mathop{\vcenter{
    \offinterlineskip\box0\box1}}}
\def\lesim{\m@thcombine<\sim}
\def\gesim{\m@thcombine>\sim}
\def\lessgtr{\m@thcombine<>}
\def\gtrless{\m@thcombine><}
\newcommand{\bra}[1]{\left\langle #1 \right|}
\newcommand{\ket}[1]{\left| #1 \right\rangle}

\newcommand{\hhat}{\hat{h}}
\newcommand{\calH}{{\cal{H}}}
\newcommand{\calHz}{{\cal{H}}_0}
\newcommand{\vecr}{\mbox{\boldmath $r$}}
\newcommand{\vecrs}{\mbox{\boldmath $r$}\sigma}
\newcommand{\vecrst}{\mbox{\boldmath $r$}\tilde{\sigma}}
\newcommand{\vecrsp}{\mbox{\boldmath $r$}'\sigma'}
\newcommand{\vecrspp}{\mbox{\boldmath $r$}''\sigma''}
\newcommand{\vecrspt}{\mbox{\boldmath $r$}'\tilde{\sigma}'}

\newcommand{\vecR}{\mbox{\boldmath $R$}}
\newcommand{\vecRz}{\mbox{\boldmath $R$}_0}
\newcommand{\vecrho}{\mbox{\boldmath $\rho$}}
\newcommand{\vecdrho}{\mbox{\boldmath $\delta\rho$}}
\newcommand{\vecv}{\mbox{\boldmath $v$}}
\newcommand{\vecone}{\mbox{\boldmath $1$}}
\newcommand{\psid}{\psi^\dag}
\newcommand{\betad}{\beta^\dag}
\newcommand{\vphi}{\varphi}

\newcommand{\htl}{\tilde{h}}
\newcommand{\vtl}{\tilde{v}}
\newcommand{\phibar}{\overline{\phi}}
\newcommand{\phibari}{\overline{\phi}_{\tilde{i}}}
\newcommand{\phibarj}{\overline{\phi}_{\tilde{j}}}
\newcommand{\rhot}{\tilde{\rho}}
\newcommand{\Vhat}{\hat{V}}
\newcommand{\calV}{{\cal{V}}}
\newcommand{\calR}{{\cal{R}}}

\newcommand{\calGz}{{\cal{G}}_{0}}
\newcommand{\calGzlj}{{\cal{G}}_{0,lj}}
\newcommand{\calGzljp}{{\cal{G}}_{0,l'j'}}

\newcommand{\Ahat}{\hat{A}}
\newcommand{\calA}{{\cal{A}}}
\newcommand{\calB}{{\cal{B}}}

\newcommand{\del}{\partial}
\newcommand{\eps}{\epsilon}
\newcommand{\Tr}{{\rm Tr}}
\newcommand{\rhohat}{\hat{\rho}}
\newcommand{\rhothat}{\hat{\tilde{\rho}}}
\newcommand{\Phat}{\hat{P}}
\newcommand{\Phatd}{\hat{P}^\dag}
\newcommand{\rsps}[2]{({\rm r}#1,+#2)}
\newcommand{\ps}{(+s)}
\newcommand{\psp}{(+s')}
\newcommand{\rs}{({\rm r}s)}
\newcommand{\rsp}{({\rm r}s')}

\begin{document}

\title{Continuum Linear Response in Coordinate Space 
Hartree-Fock-Bogoliubov Formalism for Collective Excitations
in Drip-line Nuclei}

\author{Masayuki Matsuo}

\address{
Graduate School of Science and Technology, Niigata University,
Niigata 950-2181, Japan }

\date{\today}
\maketitle

\begin{abstract}
We formulate a continuum linear response theory on the
basis of the Hartree-Fock-Bogoliubov formalism in the
coordinate space representation in order to describe 
low-lying and high-lying collective excitations which couple
to one-particle and two-particle continuum states. 
Numerical analysis is done for the neutron drip-line nucleus $^{24}$O.
A low-lying collective mode that emerges 
above the continuum threshold with large neutron strength is
analyzed. The collective state is sensitive to 
the density-dependence of the pairing. 
The present theory satisfies accurately the energy weighted sum 
rule. This is guaranteed by treating the pairing selfconsistently both in 
the static HFB and in the dynamical linear response equation.


\end{abstract}

\section{Introduction}\label{sec:intro}

Collective excitation in exotic unstable nuclei, 
especially in neutron-rich nuclei near the drip-line, is 
one of the most interesting nuclear structure issues.
The presence of the neutron halo or skin structures,
or more generally of
loosely bound neutrons and the very shallow Fermi energy
will modify characters of collective excitations 
known in stable nuclei. It is further interesting if
new kinds of collective mode emerge.
The linear response theory or the random phase approximation is
one of the powerful tools to study such issues. Since the method itself
is a general framework to describe normal modes of excitation built 
on a reference state given by mean-field
approximations\cite{Ring-Schuck}, it is advantageous to,
rather than very light drip-line nuclei  such as  $^{11}$Li, heavier
systems which will be studied in future experiments.
Previous works in this direction have 
analyzed the giant resonances and the threshold neutron 
strength in exotic nuclei with closed shell
configurations\cite{HaSaZh,FaCRPA,Colo-1}
with use of the continuum linear response theory formulated 
in the coordinate space \cite{Shlomo,Bertsch}.  
However, to explore
more systematically nuclei with open shell configurations,
the theory has to be extended so as to take into account the pairing 
correlation, which may play essential roles especially for  
low-lying collective excitations.

Indeed the pairing correlation is a key element in the study 
of nuclei near the drip-line\cite{DobHFB1,DobHFB2,Esbensen}.
A special feature of the pairing in drip-line nuclei is
that the correlation takes place not only among bound orbits in
the potential well but also in continuum states above the zero energy
threshold. The Hartree-Fock-Bogoliubov (HFB) theory formulated in 
the coordinate space representation \cite{DobHFB1,DobHFB2} 
has a great power
in this respect since it allows one to treat properly the pairing
in the continuum orbits, whereas the conventional BCS approximation 
is not suited for this purpose.
The coordinate space HFB has been applied extensively and 
clarifying some aspects of pairing effects
on the ground state properties
including the halo and the skin 
\cite{DobHFB1,DobHFB2,Sm93HFBap,Db94HFBap,Be99HFBap,Mz00HFBap,Db00HFBap,Sm88KHFB,Fa94KHFB,Fa00KHFB}.
Descriptions
of deformed exotic nuclei are also under current developments 
\cite{Te96HFBnrich1,Te97HFBnrich2,St98HFBdef,Ta00HFBdef,Ya00HFBdef}.
The coordinate space HFB has been also used together 
with the relativistic mean-field models
\cite{MeRiRMFB,Po97RMFB,La98RMFB}. 

The shallow Fermi energy in drip-line nuclei makes the
threshold for particle continuum very low.
It is therefore important to include effects 
of the continuum states in describing not only the 
ground state but also the excitation modes.
Attempts have been made to describe both the pairing correlation
and the continuum effects on collective states in the linear
response formalism. These continuum quasiparticle linear response theories
\cite{Pl88FFSth,Bo96FFSGT,Kam98FFSe1,Hagino}, 
however, rely on
the conventional BCS approximation, 
which may not be very accurate near the drip-lines. 
On the other hand, an approach is proposed to build a
quasiparticle random phase approximation (QRPA) with use of the 
canonical basis in the coordinate space HFB\cite{Engel}. 
However, the continuum effect is not precisely accounted
in this approach since the single-quasiparticle basis is discretized.
Other QRPA models using the discrete BCS quasiparticle basis 
\cite{Khan-Giai,Khan-etal,Colo-2} have a similar problem.

In the present paper, we extend the approach of Ref.\cite{Shlomo} and
formulate a new continuum 
quasiparticle linear response theory that is fully based on
the coordinate space HFB formalism, so that the theory can take into
account coupling to the continuum configurations both in describing
the pairing in the ground state and in description of collective
excitations. A novel feature is that it includes,
for the first time in the quasiparticle linear response formalism,
the configurations where excited two quasiparticles are both occupying
the continuum orbits above the threshold.
We pay a special attention
to consistency between  the treatment of the pairing
correlation in the static HFB and that 
in the linear response equation
for the collective excitations. This selfconsistency 
is quite important as we shall demonstrate in
the following. We also discuss 
some basic features of pairing effects on collective excitations,
taking as an example the quadrupole response 
in neutron rich oxygen isotopes including drip-line nucleus $^{24}$O.

The sections \S\S 2 and 3 are devoted to derivation of 
the basic equations of the linear response equations.
Numerical results for oxygen isotopes are discussed in \S 4, 
and conclusions are drawn in \S 5.

\section{Time-Dependent Hartree-Fock-Bogoliubov Equation in
Coordinate Representation and Linearization}

\subsection{TDHFB in coordinate space}

The linear response theory or the random phase approximation
can be formulated generally as a small amplitude limit of the time-dependent 
Hartree-Fock (TDHF) equation. Now in order to include the
pairing correlations,
we shall base our formulation on 
the time-dependent Hartree-Fock-Bogoliubov (TDHFB) theory.
The TDHFB equations is often formulated with use of the discrete shell model 
single-particle basis and the Thouless representation
\cite{Ring-Schuck,Blaizot-Ripka}, which however are not 
convenient to treat the continuum states. 
Thus we shall start our formulation
with  writing the basic equations of TDHFB in
the coordinate space representation. It is an extension of
the coordinate space HFB of Ref.\cite{DobHFB1} to time-dependent problems. 
We follow the notation of Ref.\cite{DobHFB1} in many aspects,
while there are some differences since we do not impose the
time-reversal symmetry assumed in Ref.\cite{DobHFB1}.

The ground state of the system $\ket{\Phi_0}$ in the HFB formalism
is a vacuum of the Bogoliubov quasiparticles. We denote the
annihilation and creation operators of the quasiparticles
by $\{ \beta_i, \beta^\dag_i\} (i=1,\cdots)$, which satisfy the
vacuum condition $\beta_i\ket{\Phi_0}=0$. 
The nucleon operators 
$\psi(\vecrs)$ and $\psid(\vecrs)$, 
where $\sigma=\pm 1/2$ denotes the spinor component,
can be expanded by the quasiparticle operators as
\begin{mathletters}\label{qpbase}
\begin{eqnarray}
\psi(\vecrs)&=&\sum_i U_i(\vecrs)\beta_i + 
                  V_i^*(\vecrs)\betad_i
            =\sum_i \vphi_{1,i}(\vecrs)\beta_i -
                  \vphi_{2,i}^*(\vecrst)\betad_i, \\
\psid(\vecrs)&=&\sum_i U_i^*(\vecrs)\betad_i + 
                  V_i(\vecrs)\beta_i
            =\sum_i \vphi_{1,i}^*(\vecrs)\betad_i - 
                  \vphi_{2,i}(\vecrst)\beta_i.
\end{eqnarray}
\end{mathletters}
Here $\vphi_{1,i}(\vecrs)$ and $\vphi_{2,i}(\vecrs)$ are the 
single-quasiparticle wave functions that satisfy the coordinate space
HFB equation \cite{DobHFB1} (see also the later description).
We do not write explicitly the isospin degrees of
freedom for simplicity although in actual applications the pairing
correlation is taken into account separately for neutrons and protons.
In the present paper we use functions
$\vphi_{1,i}(\vecrs)\equiv U_i(\vecrs)$  and
$\vphi_{2,i}(\vecrs)\equiv V_i(\vecrst)$ instead of
$U_i(\vecrs)$ and $V_i(\vecrs)$ for the convenience of notation whereas
$U_i(\vecrs)$ and $V_i(\vecrs)$  correspond more directly to 
the $U,V$ matrices in the HFB formalism in the
discrete shell model  basis \cite{Ring-Schuck}.
The symbol $\vphi(\vecrst)$ for a spinor function $\vphi(\vecrs)$
denotes $\vphi(\vecrst)\equiv(-2\sigma)\vphi(\vecr-\sigma)=
(-i\sigma_y \vphi)(\vecrs)$. For the time-reversal convention,
we employ 
$T\vphi(\vecrs)\equiv \vphi^*(\vecrst)=(-i\sigma_y \vphi^*)(\vecrs)$.
These conventions are the same as in Ref.\cite{DobHFB1}.
We also use a notation $\vphi_{\tilde{i}}(\vecrs)\equiv
\vphi^*(\vecrst)$ for the time-reversed function.

In a time dependent problem, the system is described by a
 TDHFB state vector $\ket{\Phi(t)}$. We assume that 
at the initial time $t_0$ the nucleus is in the
ground state, i.e. $\ket{\Phi(t=t_0)}=\ket{\Phi_0}$.
Time evolution of the TDHFB state vector $\ket{\Phi(t)}$ 
is governed by the time-dependent Schr{\"o}dinger equation
\begin{eqnarray}\label{TDHFB}
i\hbar{\del \over \del t}\ket{\Phi(t)}=\hhat(t)\ket{\Phi(t)}
\end{eqnarray}
with the time-dependent mean-field Hamiltonian $\hhat(t)$, which can
be expressed generally as
\begin{eqnarray}\label{Haml}
\hhat(t)= &&\int\int d\vecr d\vecr' \sum_{\sigma\sigma'}
   h(\vecrs,\vecrsp,t)\psid(\vecrs)\psi(\vecrsp)  \nonumber \\
          &&+{1\over2}\int\int d\vecr d\vecr' \sum_{\sigma\sigma'}
  \left\{\htl(\vecrs,\vecrsp,t)\psid(\vecrs)\psid(\vecrspt) 
          +\htl^*(\vecrs,\vecrsp,t)\psi(\vecrspt)\psi(\vecrs)
            \right\}  \nonumber \\
          && -\lambda\int d\vecr \sum_{\sigma}\psid(\vecrs)\psi(\vecrs),
\end{eqnarray}
where the second and the third terms represent the 
pair potential, and  $\lambda$ is the chemical potential or
the Fermi energy. 
There are symmetry properties 
$h^*(\vecrs,\vecrsp,t)=h(\vecrsp,\vecrs,t)$
and $h(\vecrs,\vecrsp,t)=h(\vecrspt,\vecrst,t)$, which stem from
the hermiticity of $\hhat(t)$ and the anti-commutation relation
of the nucleon operators.

Using the unitary operator
$\hat{U}(t)$ that describes the time evolution of $\ket{\Phi(t)}$ 
by  $\ket{\Phi(t)}=\hat{U}(t)\ket{\Phi_0}$,
the nucleon operators in the Heisenberg representation
$\psid(\vecrs t)=\hat{U}^\dagger(t)\psid(\vecrs)\hat{U}(t)$
and
$\psi(\vecrs t)=\hat{U}^\dagger(t)\psi(\vecrs)\hat{U}(t)$ are
introduced. They also have expansion
in terms of the complete single-quasiparticle basis, given by
\begin{mathletters}\label{qpbase2}
\begin{eqnarray}
\psi(\vecrs t)
            &=&\sum_i \vphi_{1,i}(\vecrs t)\beta_i -
                  \vphi_{2,i}^*(\vecrst t)\betad_i, \\
\psid(\vecrs t)
            &=&\sum_i \vphi_{1,i}^*(\vecrs t)\betad_i - 
                  \vphi_{2,i}(\vecrst t)\beta_i.
\end{eqnarray}
\end{mathletters}
The single-quasiparticle wave functions 
$\vphi_{1,i}(\vecrs t)$ and $\vphi_{2,i}(\vecrs t)$ are now time-dependent,
and they are at the initial time $t=t_0$ set 
to the ground state quasiparticle functions 
$\vphi_{1,i}(\vecrs)$ and $\vphi_{2,i}(\vecrs)$.
The field equation of motion
$i\hbar{\del\over \del t}\psi(\vecrs t)=
\left[ \psi(\vecrs t),\hhat_{\rm H}(t)\right]$
with $\hhat_{\rm H}(t)=\hat{U}^\dagger(t)\hhat(t)\hat{U}(t)$
leads to 
the time-dependent Schr{\"o}dinger equation 
for the single-quasiparticle wave functions 
$\vphi_{1,i}(\vecrs t)$ and $\vphi_{2,i}(\vecrs t)$, that is written as
\begin{equation}\label{TDHFB2}
i\hbar{\del\over \del t}
\phi_i(\vecrs t) =
\int d\vecr'\sum_{\sigma'}
\calH(\vecrs,\vecrsp,t)\phi_i(\vecrsp t), 
\end{equation}
where we have used a
$2 \times 2 $ matrix representation defined by
\begin{equation} \label{Hamlmat}
\calH(\vecrs,\vecrsp,t) \equiv 
\left(
\begin{array}{cc}
h(\vecrs,\vecrsp,t) -\lambda\delta(\vecr-\vecr')\delta_{\sigma\sigma'} & 
                                    \htl(\vecrs,\vecrsp,t) \\
\htl^*(\vecrst,\vecrspt,t) & 
  -h^*(\vecrst,\vecrspt,t)+\lambda\delta(\vecr-\vecr')\delta_{\sigma\sigma'}
\end{array}
\right),
\end{equation}
and
\begin{equation} \label{spinor}
\phi_i(\vecrs t) \equiv 
\left(
\begin{array}{c}
\vphi_{1,i}(\vecrs t) \\
\vphi_{2,i}(\vecrs t)
\end{array}
\right).
\end{equation}
Note also that the same time-dependent single-quasiparticle
equation (\ref{TDHFB2}) holds for $\phibari(\vecrs t)$ defined by
\begin{equation} \label{conjugate}
\phibari(\vecrs t) \equiv 
\left(
\begin{array}{c}
-\vphi^*_{2,i}(\vecrst t) \\
\vphi^*_{1,i}(\vecrst t)
\end{array}
\right) =
\left(
\begin{array}{cc}
0 & -1 \\
1 & 0
\end{array}
\right)
\phi_{\tilde{i}}(\vecrs t).
\end{equation}
The bar implies the operation of
$\left(
\begin{array}{cc}
0 & -1 \\
1 & 0
\end{array}
\right)$. With these notations, Eq.(\ref{qpbase2}) is written in a compact form
\begin{equation}\label{field}
\left(
\begin{array}{c}
\psi(\vecrs t) \\
\psid(\vecrst t)
\end{array}
\right) =\sum_i \phi_i(\vecrs t)\beta_i +\phibari(\vecrs t)\betad_i.
\end{equation}
A conjugate pair is formed by 
$\phi_i(\vecrs t)$ and $\phibari(\vecrs t)$.
The orthonormality condition and the completeness for the 
single-quasiparticle states are expressed as
\begin{mathletters}\label{complete}
\begin{eqnarray}
&&\int d\vecr \sum_\sigma \phi_i^\dag(\vecrs t)\phi_j(\vecrs t)=
\int d\vecr \sum_\sigma \phibari^\dag(\vecrs t)\phibarj(\vecrs t)
=\delta_{ij}, \\
&&\int d\vecr \sum_\sigma \phi_i^\dag(\vecrs t)\phibarj(\vecrs t)=0,\\
&&\sum_i \phi_i(\vecrs t)\phi_i^\dag(\vecrsp t)
   +\phibari(\vecrs t)\phibari^\dag(\vecrsp t) 
= \delta(\vecr-\vecr')\delta_{\sigma\sigma'}
\left(
\begin{array}{cc}
1 & 0 \\
0 & 1 
\end{array}
\right) .
\end{eqnarray}
\end{mathletters}

It is convenient to use the time-dependent density matrices in
describing the evolution of the TDHFB state vector. We define 
the normal and the abnormal density (pair density) matrices by 
\begin{mathletters}\label{density}
\begin{eqnarray}
&&\rho(\vecrs,\vecrsp,t) \equiv 
     \bra{\Phi(t)}\psid(\vecrsp)\psi(\vecrs)\ket{\Phi(t)}=
     \bra{\Phi_0}\psid(\vecrsp t)\psi(\vecrs t)\ket{\Phi_0}, \\
&&\rhot(\vecrs,\vecrsp,t) \equiv 
     \bra{\Phi(t)}\psi(\vecrspt)\psi(\vecrs)\ket{\Phi(t)}=
     \bra{\Phi_0}\psi(\vecrspt t)\psi(\vecrs t)\ket{\Phi_0},
\end{eqnarray}
\end{mathletters}
respectively. These definitions agree with those in Ref.\cite{DobHFB1}
if the state vector time-independent $\ket{\Phi(t)}=\ket{\Phi_0}$
and the HFB ground state is time-even $T\ket{\Phi_0}=\ket{\Phi_0}$,
although in this paper such assumption is not made.
Also our definition of $h(\vecrs,\vecrsp,t)$ and $\htl(\vecrs,\vecrsp,t)$
reduces to those in  Ref.\cite{DobHFB1} with the same conditions.
The two density matrices are combined in a  
generalized density matrix $\calR$ as 
\begin{eqnarray}\label{gendens}
\calR(\vecrs,\vecrsp,t) &&\equiv
\left(
\begin{array}{cc}
\bra{\Phi(t)}\psid(\vecrsp)\psi(\vecrs)\ket{\Phi(t)} &
\bra{\Phi(t)}\psi(\vecrspt)\psi(\vecrs)\ket{\Phi(t)} \\
\bra{\Phi(t)}\psid(\vecrsp)\psid(\vecrst)\ket{\Phi(t)} &
\bra{\Phi(t)}\psi(\vecrspt)\psid(\vecrst)\ket{\Phi(t)} \\
\end{array}
\right) \nonumber \\
&& =
\left(
\begin{array}{cc}
\rho(\vecrs,\vecrsp,t) & \rhot(\vecrs,\vecrsp,t) \\
\rhot^*(\vecrst,\vecrspt,t) & 
     \delta(\vecr-\vecr')\delta_{\sigma\sigma'} - \rho^*(\vecrst,\vecrspt,t) 
\end{array}
\right).
\end{eqnarray}
The generalized density matrix can also be expressed in terms
of the time-dependent single-quasiparticle wave functions as
\begin{equation}
\calR(\vecrs,\vecrsp,t) = \sum_i \phibari(\vecrs t)\phibari^\dag(\vecrsp t).
\end{equation}

\subsection{Linearization}

Let us consider a TDHFB state vector $\ket{\Phi(t)}$ which
fluctuates around the HFB ground state $\ket{\Phi_0}$ under 
perturbation of an external field.  The external perturbation
induces also fluctuation in the selfconsistent mean-field.
We now write the time-dependent mean-field Hamiltonian
as $\hhat(t)=\hhat_0 + \Vhat(t)$ where $\hhat_0$ denotes the
static selfconsistent HFB mean-field Hamiltonian for the ground state while
the fluctuating field $\Vhat(t)$ contains both 
the external field $\Vhat^{\rm ext}(t)$ 
and the induced field $\Vhat^{\rm ind}(t)$. 
With presence of the pair correlation, the fluctuating
field $\Vhat(t)$ is 
a generalized one-body operator including pair creation and
annihilation, which can be expressed as
\begin{eqnarray}\label{vfield}
&\Vhat(t)=& \int\int d\vecr d\vecr' \sum_{\sigma\sigma'}
     \left\{    
  v(\vecrs,\vecrsp,t)\psid(\vecrs)\psi(\vecrsp) \right. \nonumber \\
        &&\left. +{1\over2}\vtl(\vecrs,\vecrsp,t)\psid(\vecrs)\psid(\vecrspt)
          +{1\over2}\vtl^*(\vecrs,\vecrsp,t)\psi(\vecrspt)\psi(\vecrsp)
\right\},
\end{eqnarray}
where we assume $\Vhat(t)$ is hermite.
To describe response of 
the time-dependent single-quasiparticle wave function 
$\phi_i(\vecrs t)$ to the perturbation, we expand
it up to the linear order as
\begin{equation}\label{lin1}
\phi_i(\vecrs t)=e^{-iE_i(t-t_0)/\hbar}\left(
\phi_i(\vecrs)+\delta\phi_i(\vecrs t)\right).
\end{equation}
Here $\phi_i(\vecrs)$ and $E_i$ denote the quasiparticle
functions and the quasiparticle excitation energy defined as 
a solution of the static HFB equation for the ground state,
\begin{equation}\label{grHFB}
\calHz \phi_i(\vecrs) = E_i \phi_i(\vecrs),
\end{equation}
where $\calHz$ is the $2 \times 2$ matrix representation of 
the static HFB mean-field Hamiltonian $\hhat_0$.

An useful tool in describing the linear response is
the single-particle Green function, which in our case is the
single-quasiparticle Green function defined for the static HFB mean-field
Hamiltonian $\hhat_0$. Because of the pairing correlation,
we need both the normal and abnormal Green functions
\begin{mathletters}\label{green1}
\begin{eqnarray}
\label{green1a}
&&G_0(\vecrs t,\vecrsp t')=-i\theta(t-t')
\bra{\Phi_0}\{\psi(\vecrs t),\psid(\vecrsp t')\}\ket{\Phi_0}, \\
\label{green1b}
&& F_0(\vecrs t,\vecrsp t')=-i\theta(t-t')
\bra{\Phi_0}\{\psid(\vecrst t),\psid(\vecrsp t')\}\ket{\Phi_0}.
\end{eqnarray}
\end{mathletters}
Here the nucleon operators 
$\psid(\vecrs t)=\hat{U}_0^\dag(t)
\psid(\vecrs)\hat{U}_0(t)$ and
$\psi(\vecrs t)=\hat{U}_0^\dag(t)
\psi(\vecrs)\hat{U}_0(t)$ with 
$\hat{U}_0(t)=e^{-i(t-t_0)\hhat_0/\hbar}$
evolve in time under the
static HFB mean-field Hamiltonian $\hhat_0$.
More useful is the single-quasiparticle Green function 
in the $2\times 2$ 
matrix representation \cite{Belyaev,Fetter-Walecka} defined by
\begin{equation}\label{green2}
\calGz(\vecrs t, \vecrsp t')=-i\theta(t-t')
\left(
\begin{array}{cc}
\bra{\Phi_0}\{\psi(\vecrs t),\psid(\vecrsp t')\}\ket{\Phi_0} &
\bra{\Phi_0}\{\psi(\vecrs t),\psi(\vecrspt t')\}\ket{\Phi_0} \\
\bra{\Phi_0}\{\psid(\vecrst t),\psid(\vecrsp t')\}\ket{\Phi_0} &
\bra{\Phi_0}\{\psid(\vecrst t),\psi(\vecrspt t')\}\ket{\Phi_0} 
\end{array}
\right).
\end{equation}
The normal and abnormal Green
functions $G_0$ and $F_0$ are contained in the 11 and 21 components, 
respectively. 
In the present paper we use the {\it retarded} functions in
stead of the {\it causal} or {\it Feynman} functions.
We call $\calGz$ the HFB Green function.
The Fourier transform of the HFB 
Green function $\calGz(\vecrs t, \vecrsp t')$
is given by
$\calGz(\vecrs,\vecrsp,E+i\eps)=\left(E+i\eps-\calHz\right)^{-1}$
where the infinitesimal constant $\eps \rightarrow +0$ exhibits the
causality. In actual numerical calculations, $\eps$ is fixed to
a small but finite number. Note that 
the HFB Green function can be constructed in
different ways. If we adopt the spectral representation, it is
expressed as
\begin{equation}\label{green3}
\calGz(\vecrs,\vecrsp,E)=
   \sum_i{\phi_i(\vecrs)\phi_i^\dag(\vecrsp)\over E-E_i}
+  {\phibari(\vecrs)\phibari^\dag(\vecrsp)\over E+E_i}.
\end{equation}
Another form which is useful to treat
the continuum states is given in Ref.\cite{Belyaev}, which we
utilize in the following.

The Fourier transform of the linear response $\delta\phi_i$ of
the single-quasiparticle wave function is now expressed as
\begin{equation}\label{lin3}
 \delta\phi_i(\vecrs,\omega)=\int\int d\vecr'd\vecr''\sum_{\sigma'\sigma''}
\calGz(\vecrs,\vecrsp,\hbar\omega+i\eps+E_i)
\calV(\vecrsp,\vecrspp,\omega)\phi_i(\vecrspp)
\end{equation}
in terms of the HFB Green function. Here $\calV$ is the matrix representation 
of the fluctuating field $\Vhat(t)$, 
which is defined in the frequency domain by
\begin{equation}
\calV(\vecrs,\vecrsp,\omega)=\left(
\begin{array}{cc}
v(\vecrs,\vecrsp,\omega) & \vtl(\vecrs,\vecrsp,\omega) \\
\vtl^*(\vecrst,\vecrspt,\omega) & -v^*(\vecrst,\vecrspt,\omega)
\end{array}
\right).
\end{equation}
For the conjugate wave function  $\phibari$ holds the same equation 
except that the energy argument of the HFB Green function is replaced
by $\hbar\omega+i\eps-E_i$.

Using the above results, the linear response in the density matrix
$\calR(\omega)=\calR^{(0)}+\delta\calR(\omega)$ 
is obtained as
\begin{eqnarray}\label{lindens}
\delta\calR(\vecrs,\vecrsp,\omega)=
\int\int d\vecr_1 d\vecr_2 \sum_{\sigma_1\sigma_2} \sum_i 
&&\left\{ \calGz(\vecrs,\vecr_1\sigma_1,-E_i+\hbar\omega+i\eps)
\calV(\vecr_1\sigma_1,\vecr_2\sigma_2,\omega)\phibari(\vecr_2\sigma_2)
\phibari^{\dag}(\vecrsp) \right. \nonumber \\
&& \left. +\phibari(\vecrs)
\phibari^{\dag}(\vecr_1\sigma_1)
\calV(\vecr_1\sigma_1,\vecr_2\sigma_2,\omega)
\calGz(\vecr_2\sigma_2,\vecrsp,-E_i-\hbar\omega-i\eps) \right\}.
\end{eqnarray}
Expectation value $A(t)=\bra{\Phi(t)}\Ahat\ket{\Phi(t)}$ 
of a one-body operator 
\begin{eqnarray}\label{lina1}
\Ahat= {1\over 2}\int\int d\vecr d\vecr' \sum_{\sigma\sigma'}
   && \left\{ \calA_{11}(\vecrs,\vecrsp)\psid(\vecrs)\psi(\vecrsp)
  +\calA_{22}(\vecrs,\vecrsp)\psi(\vecrst)\psid(\vecrspt) \right.\nonumber \\
   &&\left. +\calA_{12}(\vecrs,\vecrsp)\psid(\vecrs)\psid(\vecrspt)
  +\calA_{21}(\vecrs,\vecrsp)\psi(\vecrspt)\psi(\vecrsp) \right\}
\end{eqnarray}
is expressed in terms of the 
generalized density matrix as
\begin{equation}
A(t)=\int\int d\vecr d\vecr'\sum_{\sigma\sigma'} 
\Tr \calA(\vecrs,\vecrsp)\calR(\vecrsp,\vecrs,t)
\end{equation}
with
\begin{equation}
\calA(\vecrs,\vecrsp)=\left(
\begin{array}{cc}
\calA_{11}(\vecrs,\vecrsp) &  \calA_{12}(\vecrs,\vecrsp) \\
\calA_{21}(\vecrs,\vecrsp) &  \calA_{22}(\vecrs,\vecrsp) 
\end{array}
\right),
\end{equation}
where $\Tr$ denotes the trace with respect to the 
$2\times 2$ matrix.
Linear response of $A(t)$ in the Fourier representation is
given by
\begin{eqnarray}\label{lina2}
\delta A(\omega)
&&=\int\cdots\int d\vecr_1 d\vecr_2 d\vecr_3 d\vecr_4
\sum_{\sigma_1\sigma_2\sigma_3\sigma_4} \sum_i \nonumber\\
&&\left\{ \Tr \calA(\vecr_1\sigma_1,\vecr_2\sigma_2)
    \calGz(\vecr_2\sigma_2,\vecr_3\sigma_3,-E_i+\hbar\omega+i\eps)
    \calV(\vecr_3\sigma_3,\vecr_4\sigma_4,\omega)
    \phibari(\vecr_4\sigma_4)\phibari^\dag(\vecr_1\sigma_1) \right.\nonumber\\
&&\left. +\Tr \calA(\vecr_1\sigma_1,\vecr_2\sigma_2)
    \phibari(\vecr_2\sigma_2)\phibari^\dag(\vecr_3\sigma_3)
    \calV(\vecr_3\sigma_3,\vecr_4\sigma_4,\omega)
    \calGz(\vecr_4\sigma_4,\vecr_1\sigma_1,-E_i-\hbar\omega-i\eps) \right\}.
\end{eqnarray}

\subsection{Unperturbed response functions for local densities}

In the following we assume that the fluctuating 
field $\Vhat(t)$ including both the external and the induced fields is
expressed in terms of the spin-independent local density
\begin{equation}
\rhohat(\vecr)=\sum_\sigma \psid(\vecrs)\psi(\vecrs)
\end{equation}
and the local pair densities
\begin{mathletters}
\begin{eqnarray}
&\Phatd(\vecr)&={1\over2}\sum_\sigma \psid(\vecrs)\psid(\vecrst), \\
&\Phat(\vecr)&={1\over2}\sum_\sigma \psi(\vecrst)\psi(\vecrs).
\end{eqnarray}
\end{mathletters}
The above assumption is valid when the residual nuclear interaction is
a zero-range force. For the convenience of notation, we
use a symmetrized form for the pair densities, defined by
\begin{mathletters}
\begin{eqnarray}
&\rhothat_+(\vecr)&=\Phatd(\vecr)+\Phat(\vecr), \\
&\rhothat_-(\vecr)&=\Phatd(\vecr)-\Phat(\vecr). 
\end{eqnarray}
\end{mathletters}
The fluctuating field is expressed as
\begin{equation}\label{Vlocal}
\Vhat(t)=\int d\vecr \left( v_0(\vecr,t)\rhohat(\vecr)
            +v_+(\vecr,t)\rhothat_+(\vecr)
            +v_-(\vecr,t)\rhothat_-(\vecr) \right).
\end{equation}

We consider now  linear response in expectation values
of the local density operators 
\begin{mathletters}
\begin{eqnarray}
&\rho(\vecr,t)&\equiv\bra{\Phi(t)}\rhohat(\vecr)\ket{\Phi(t)}
=\sum_\sigma\rho(\vecrs,\vecrs,t), \\
&\rhot_+(\vecr,t)&\equiv\bra{\Phi(t)}\rhothat_+(\vecr)\ket{\Phi(t)}
={1\over2}\sum_\sigma
   \left(\rhot^*(\vecrs,\vecrs,t)+\rhot(\vecrs,\vecrs,t)\right), \\
&\rhot_-(\vecr,t)&\equiv\bra{\Phi(t)}\rhothat_-(\vecr)\ket{\Phi(t)}
={1\over2}\sum_\sigma
   \left(\rhot^*(\vecrs,\vecrs,t)-\rhot(\vecrs,\vecrs,t)\right),
\end{eqnarray}
\end{mathletters}
under the influence of the fluctuating field $\Vhat(t)$ and 
the static HFB mean-field Hamiltonian $\hhat_0$. 
The relation between  $\Vhat(t)$ and the density linear responses
is expressed as 
\begin{equation}\label{linresp1}
\delta\rho_\alpha(\vecr,t)=\int_{t_0}^{t}dt'\int d\vecr'
\sum_\beta R_{0}^{\alpha\beta}(\vecr t,\vecr' t')v_\beta(\vecr',t')
\end{equation}
by means of the unperturbed response function 
$R_{0}^{\alpha\beta}(\vecr t,\vecr' t')$, whose formal expression is
\cite{Fetter-Walecka}
\begin{equation}\label{respf}
R_{0}^{\alpha\beta}(\vecr t,\vecr' t') = -i\theta(t-t')\bra{\Phi_0}
\left[\hat{U}_0^\dagger(t)\rhohat_\alpha(\vecr)\hat{U}_0(t),
\hat{U}_0^\dagger(t')\rhohat_\beta(\vecr')\hat{U}_0(t')\right]\ket{\Phi_0}.
\end{equation}
Here $\rho_\alpha(\vecr,t)$ represents the
three kinds of density 
$\rho(\vecr,t),\rhot_+(\vecr,t)$, and $\rhot_-(\vecr,t)$. 
We also denote $\rhohat_\alpha(\vecr)=
\rhohat(\vecr),\rhothat_+(\vecr), \rhothat_-(\vecr)$, and
$v_\alpha(\vecr,t)=v_0(\vecr,t), v_+(\vecr,t), v_-(\vecr,t)$, respectively.
With use of Eq.(\ref{lina2}), the Fourier transform of 
the unperturbed response function is obtained as
\begin{eqnarray}\label{uresp1}
R_{0}^{\alpha\beta}(\vecr,\vecr',\omega) ={1\over2}\sum_i \sum_{\sigma\sigma'}
&&\left\{
 \phibari^{\dag}(\vecrs)\calA\calGz(\vecrs,\vecrsp,-E_i+\hbar\omega+i\eps)
\calB\phibari(\vecrsp) \right. \nonumber\\
&& \left. + 
\phibari^{\dag}(\vecrsp)\calB\calGz(\vecrsp,\vecrs,-E_i-\hbar\omega-i\eps)
\calA\phibari(\vecrs)\right\} 
\end{eqnarray}
in terms of the HFB Green function $\calGz$.
Here $\calA$ and $\calB$ are $2 \times 2$ matrices which
correspond to the normal and abnormal densities; 
$\calA=\left(\begin{array}{cc}2 & 0 \\ 0&0 \end{array}\right),
\left(\begin{array}{cc}1 & 0 \\ 0&1 \end{array}\right)$,
and $\left(\begin{array}{cc}0 & 1 \\ 1&0 \end{array}\right)$ for
$\rhohat_\alpha(\vecr)=\rhohat(\vecr),\rhothat_+(\vecr)$, and 
$\rhothat_-(\vecr)$,
respectively. 
$\calB=\left(\begin{array}{cc}1 & 0 \\ 0&1 \end{array}\right)$ for
$\rhohat_\beta(\vecr)=\rhohat(\vecr)$, and  $\calB$ is the same as $\calA$
for $\rhohat_\beta(\vecr)=\rhothat_+(\vecr), \rhothat_-(\vecr)$.

If we insert Eq.(\ref{green3}) 
to the above equation, the unperturbed response function reduces
to a spectral representation in a standard form
\begin{eqnarray}\label{uresp2}
R_{0}^{\alpha\beta}(\vecr,\vecr',\omega) ={1\over2}\sum_{ij}
&& \left\{  
\bra{0}\rhohat_\alpha(\vecr)\ket{ij}\bra{ij}\rhohat_\beta(\vecr')\ket{0}
{1 \over \hbar\omega +i\eps-E_i-E_j } \right. \nonumber\\
&& \left. -
\bra{0}\rhohat_\beta(\vecr')\ket{ij}\bra{ij}\rhohat_\alpha(\vecr)\ket{0}
{1 \over \hbar\omega +i\eps+E_i+E_j}
\right\} ,
\end{eqnarray}
where
\begin{mathletters}
\begin{eqnarray}
&&\bra{ij}\rhohat_\alpha(\vecr)\ket{0}
  =\sum_\sigma \phi_i^\dag(\vecrs)\calA\phibarj(\vecrs), \\
&&\bra{0}\rhohat_\alpha(\vecr)\ket{ij}
  =\sum_\sigma \phibarj^\dag(\vecrs)\calA\phi_i(\vecrs).
\end{eqnarray}
\end{mathletters}
This can also be derived directly from Eq.(\ref{respf}). 
This expression, however,
is not convenient to treat the continuum states.

\section{Correlated Linear Response with Continuum States}

\subsection{Integral representation}

If the nucleus is put in the space of infinite volume, the 
spectrum of HFB single-quasiparticle states become continuum for the
quasiparticle excitation energy $E$ which exceeds the one-particle separation
energy 
$S_{1}=|\lambda|$ \cite{DobHFB1,Belyaev}. 
Accordingly, the single-quasiparticle HFB Green function $\calGz$ also 
exhibits the continuum spectrum. Furthermore,  
the HFB Green function at the continuum
energy is required to satisfy the boundary condition of outgoing
wave in the exterior region of the nucleus\cite{Belyaev}, which is
a suitable boundary condition for the continuum states.
Let's us implement these features in the response functions.

\begin{figure}[t]
\centerline{\psfig{figure=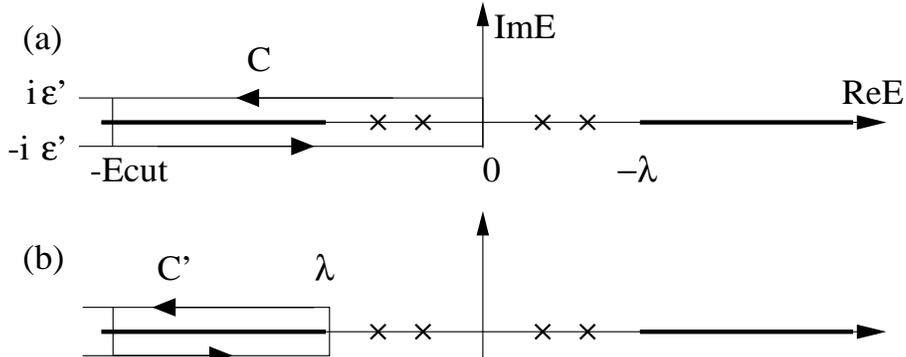,width=12cm}}
\caption{Contours $C$ and $C'$ in the integral representation of the
response function. The crosses represent the poles at
$E=\pm E_i$ corresponding to the bound quasiparticle states. The
thick lines are the branch cuts.}
\end{figure}

This task is accomplished in two steps. Firstly, we
handle the continuity of the HFB quasiparticle spectrum.
For this purpose, we rewrite the summation over the HFB quasiparticles
in Eq.(\ref{uresp1}) 
with use of an 
integral representation. Namely, a summation 
$\sum_i f(-E_i)\phibari(\vecr)\phibari^\dag(\vecrs)$ 
is replaced by a contour integral in the complex $E$ plane
as
\begin{equation}
\sum_i f(-E_i)\phibari(\vecr)\phibari^\dag(\vecrs)=
{1\over2\pi i}\int_C dE f(E)\calGz(E),
\end{equation}
in terms of the HFB Green function $\calGz$, 
Eq.(\ref{green3}), 
which has poles at $E=-E_i$ on the negative energy axis.
The contour has to be chosen so that it encloses these poles,
and at the same time, it must avoid the poles associated
with $\calGz(E+\hbar\omega+i\eps)$ and $\calGz(E-\hbar\omega-i\eps)$
that account for $\calGz(-E_i+\hbar\omega+i\eps)$ and
$\calGz(-E_i-\hbar\omega-i\eps)$ in Eq.(\ref{uresp1}). 
A contour $C$ satisfying this
requirement is shown in Fig.1(a), where the constant $\eps'$ 
must satisfy $\eps > \eps' >0$. 
Consequently, the unperturbed response function
is expressed as
\begin{eqnarray}\label{uresp3}
R_{0}^{\alpha\beta}(\vecr,\vecr',\omega) =
{1\over 4\pi i}\int_C dE \sum_{\sigma\sigma'}
&&\left\{ 
\Tr\calA\calGz(\vecrs,\vecrsp,E+\hbar\omega+i\eps)
        \calB\calGz(\vecrsp,\vecrs,E)  \right. \nonumber\\
&&\left. +\Tr\calA\calGz(\vecrs,\vecrsp,E)
        \calB\calGz(\vecrsp,\vecrs,E-\hbar\omega-i\eps) \right\}.
\end{eqnarray}
The contour is the same as those adopted in
Refs.\cite{Belyaev,Fa00KHFB} except the condition for $\eps'$.

For spherically symmetric systems, the partial wave
expansion can be applied as
\begin{equation}
\phi(\vecrs)=Y_{ljm}(\hat{\vecr}\sigma) {1\over r}\phi_{lj}(r), \ \ \ 
\phi_{lj}(r)=\left(
\begin{array}{c} \vphi_{1,lj}(r) \\ \vphi_{2,lj}(r) \end{array}
\right),
\end{equation}
\begin{equation}
\calGz(\vecrs,\vecrsp,E)=\sum_{ljm}
 Y_{ljm}(\hat{\vecr}\sigma) {1\over rr'}\calGzlj(r,r',E)
Y^*_{ljm}(\hat{\vecr'}\sigma'),
\end{equation}
\begin{equation}
R_{0}^{\alpha\beta}(\vecr,\vecr',\omega)=\sum_{LM}
 Y_{LM}(\hat{\vecr}) {1\over r^2{r'}^2}
R_{0,L}^{\alpha\beta}(r,r',\omega)
Y^*_{LM}(\hat{\vecr'}),
\end{equation}
with $Y_{ljm}(\hat{\vecr}\sigma)$ being the spin spherical harmonics. 
The unperturbed response function with the multipole $L$ is then given by
\begin{eqnarray}\label{uresp5}
R_{0,L}^{\alpha\beta}(r,r',\omega) =
{1\over 4\pi i}\int_C dE \sum_{lj,l'j'}
{\left<l'j'\right\|Y_L\left\|lj\right>^2 \over 2L+1} 
&& \left\{
\Tr\calA\calGzljp(r,r',E+\hbar\omega+i\eps)
\calB\calGzlj(r',r,E) \right. \nonumber\\
&& \left. + 
\Tr\calA\calGzlj(r,r',E)\calB\calGzljp(r',r,E-\hbar\omega-i\eps)
 \right\}.
\end{eqnarray}
If we treat separately the discrete and the continuum parts of the
HFB spectrum, the following equivalent expression is obtained
\begin{eqnarray}\label{uresp6}
R_{0,L}^{\alpha\beta}(r,r',\omega) =
&&{1\over2} \sum_{lj,l'j'}\sum'_n  
{\left<l'j'\right\|Y_L\left\|lj\right>^2 \over 2L+1} \left\{
\phibar_{nlj}^{T}(r)\calA\calGzljp(r,r',-E_{nlj}+\hbar\omega+i\eps)
\calB\phibar_{nlj}(r')  \right. \nonumber\\
&&\hspace{30mm} \left. + 
\phibar_{nlj}^{T}(r')\calB\calGzljp(r',r,-E_{nlj}-\hbar\omega-i\eps)
\calA\phibar_{nlj}(r) \right\} \nonumber\\
&& +{1\over 4\pi i}\int_{C'} dE \sum_{lj,l'j'} 
{\left<l'j'\right\|Y_L\left\|lj\right>^2 \over 2L+1} 
 \left\{
\Tr\calA\calGzljp(r,r',E+\hbar\omega+i\eps)
\calB\calGzlj(r',r,E) \right. \nonumber\\
&& \hspace{30mm} \left. + 
\Tr\calA\calGzlj(r,r',E)\calB\calGzljp(r',r,E-\hbar\omega-i\eps)
 \right\},
\end{eqnarray}
where the summation $\sum'_n$ runs only over the bound single-quasiparticle
states with discrete spectrum $E_{nlj} < |\lambda|$, while the contour
$C'$ in the complex plane encloses only the continuum part, as shown
in Fig.1(b).

Having the integral representation of the response functions,
we impose on the 
single-quasiparticle Green function the boundary condition
of out-going wave in the exterior region. The HFB Green
function satisfying this requirement is given in
Ref.\cite{Belyaev,Fa00KHFB}, which can
be applied to spherically symmetric systems with local
potential. Namely,
the HFB Green function is constructed \cite{Belyaev,Fa00KHFB}as
\begin{equation}\label{greenex}
\calGzlj(r,r',E)=\sum_{s,s'=1,2}
c_{lj}^{ss'}(E)\left(
\theta(r-r')\phi_{lj}^{\ps}(r,E){\phi_{lj}^{\rsp T}}(r',E)
+\theta(r'-r)\phi_{lj}^{\rsp}(r,E){\phi_{lj}^{\ps T}}(r',E)
\right).
\end{equation}
Here
$\phi_{lj}^{\rs}(r,E) (s=1,2)$ are 
two independent solutions, regular at the origin $r=0$, of
the radial HFB equation
\begin{equation}
\left(
\begin{array}{cc}
{-\hbar^2 \over 2m}{d^2 \over dr^2}+U_{lj}(r)-\lambda & \Delta(r) \\
\Delta(r) & {\hbar^2 \over 2m}{d^2 \over dr^2}-U_{lj}(r)+\lambda 
\end{array}
\right)
\phi_{lj}(r,E) = E \phi_{lj}(r,E),
\end{equation}
while $\phi_{lj}^{\ps}(r,E) (s=1,2)$ are two independent 
solutions that satisfy the 
boundary condition at $r\rightarrow \infty$ 
\begin{equation}
\phi_{lj}^{(+1)}(r,E) \rightarrow 
\left( \begin{array}{c} e^{ik_+(E)r} \\ 0\end{array}\right), \ \ \ \
\phi_{lj}^{(+2)}(r,E) \rightarrow 
\left( \begin{array}{c} 0 \\ e^{ik_-(E)r}\end{array}\right).
\end{equation}
Here $k_{\pm}(E)=\sqrt{2m(\lambda\pm E)}/\hbar$ and 
the branch cuts are chosen so that ${\rm Im}k_\pm >0$ is satisfied.
Thus the nucleons at $r\rightarrow \infty$ is out-going for
$E+i\eps$ at the continuum energy $E>|\lambda|$ \cite{Belyaev,Fa00KHFB}.
The coefficients $c_{lj}^{ss'}(E)$ are expressed in terms of the
Wronskians as
\begin{equation}
\left(
\begin{array}{cc}
c_{lj}^{11} & c_{lj}^{12} \\ 
c_{lj}^{21} & c_{lj}^{22} 
\end{array}
\right)
=
\left(
\begin{array}{cc}
w_{lj}\rsps{1}{1} & w_{lj}\rsps{1}{2} \\ 
w_{lj}\rsps{2}{1} & w_{lj}\rsps{2}{2}
\end{array}
\right)^{-1}
\end{equation}
with
\begin{equation}
w_{lj}\rsps{s}{s'}={\hbar^2 \over 2m}\left(
\vphi_{1,lj}^{\rs}(r){d\over dr}\vphi_{1,lj}^{\psp}(r)
-\vphi_{1,lj}^{\psp}(r){d\over dr}\vphi_{1,lj}^{\rs}(r)
-\vphi_{2,lj}^{\rs}(r){d\over dr}\vphi_{2,lj}^{\psp}(r)
+\vphi_{2,lj}^{\psp}(r){d\over dr}\vphi_{2,lj}^{\rs}(r)
\right).
\end{equation}

\subsection{RPA response functions}

Let us now describe linear response of the system 
by including correlation effects caused by 
the residual interactions among nucleons. The correlation
brings about the induced field, which can be described as a 
fluctuating part of
the selfconsistent mean-field associated with the TDHFB state vector
$\ket{\Phi(t)}$. Generally, the expectation value of the total energy for
$\ket{\Phi(t)}$ is a functional $E[\calR]=E[\rho,\rhot,\rhot^*]$ of 
the generalized density matrix $\calR$, or  $\rho,\rhot$, and $\rhot^*$. 
The selfconsistent mean-fields $h$ and $\htl$ in Eq.(\ref{Haml})
are then defined by a derivative of the functional with
respect to the densities,
\begin{mathletters}
\begin{eqnarray}
&&h(\vecrs,\vecrsp,t)={\del E \over \del \rho(\vecrsp,\vecrs,t)}, \\
&&\htl(\vecrs,\vecrsp,t)=2{\del E \over \del \rhot^*(\vecrsp,\vecrs,t)}.
\end{eqnarray}
\end{mathletters}
If we assume zero-range effective interactions, the energy functional
is expressed in terms of the local spin-independent densities
$\rho_\alpha(\vecr,t)=\rho(\vecr,t),\rhot_+(\vecr,t)$ and 
$\rhot_-(\vecr,t)$. In this case, the selfconsistent mean-field 
is expressed in the same form as Eq.(\ref{Vlocal}) 
with the field functions given by 
\begin{equation}
v_\alpha(\vecr,t)={\del E \over \del \rho_\alpha(\vecr,t)}.
\end{equation}
Accordingly, the induced fields are
\begin{equation}
v_\alpha^{ind}(\vecr,t)=\sum_\beta 
      \left({\del v_\alpha \over \del \rho_\beta}\right)_{\rm gs}
 \hspace{-10pt}(\vecr)
             \delta\rho_\beta(\vecr,t)
  =\sum_{\beta}
\left({\del^2 E \over \del \rho_\alpha\del\rho_\beta}\right)_{\rm gs}
 \hspace{-10pt}(\vecr)
             \delta\rho_\beta(\vecr,t).
\end{equation}
Inserting $v_\alpha=v_\alpha^{ind}+v_\alpha^{ext}$ to Eq.(\ref{linresp1}), 
we obtain the equation for the density linear response 
$\delta\rho_\alpha(\vecr,t)=\delta\rho(\vecr,t),
\delta\rhot_+(\vecr,t)$, and $\delta\rhot_-(\vecr,t)$, with the
correlation effect taken into account. That is,
\begin{equation}\label{respeq1}
\delta\rho_\alpha(\vecr,\omega)=\int d\vecr \sum_\beta
R_0^{\alpha\beta}(\vecr,\vecr',\omega)\left(
\sum_\gamma{\del v_\beta \over \del \rho_\gamma}(\vecr')
          \delta\rho_\gamma(\vecr',\omega)
+v^{ext}_\beta(\vecr',\omega)\right).
\end{equation}
This linear response equation is similar to the one in 
the continuum linear response theory for unpaired systems
\cite{Shlomo}, but we here take into account the fluctuations in
the pair densities.
It is possible to write the above equation as
\begin{equation}
\vecdrho(\omega)=\vecRz(\omega){\del\vecv\over\del\vecrho}\vecdrho(\omega)
+ \vecRz(\omega)\vecv^{ext},
\end{equation}
where the three kinds of density fluctuations are
represented as a single extended vector $\vecdrho(\omega)=
(\delta\rho(\vecr,\omega),\delta\rhot_+(\vecr,\omega),
\delta\rhot_-(\vecr,\omega))^T$ with three components, and
the same representation applies to 
the external fields $\vecv^{ext}$. 
We consider both neutrons and protons in the actual application.
Then the density response $\vecdrho$ have six components 
(three for each isospin).
The solution of the linear response equation is 
expressed as
\begin{equation}
\vecdrho(\omega)=\left(\vecone-\vecRz(\omega)
{\del\vecv\over\del\vecrho}\right)^{-1}
\vecRz(\omega)\vecv^{ext} \equiv \vecR(\omega)\vecv^{ext}
\end{equation}
or
\begin{equation}
\delta\rho_\alpha(\vecr,\omega)=\int d\vecr' \sum_\beta
R^{\alpha\beta}(\vecr,\vecr',\omega)v_\beta^{ext}(\vecr'),
\end{equation}
where $\vecR(\omega)=\left( R^{\alpha\beta}(\vecr,\vecr',\omega) \right)$
is the correlated response function. This is nothing but the
response function in the random phase approximation (RPA). 

The density response of the spherical system 
to an external field with the multipolarity $L$ is given by
\begin{eqnarray}
&&v_\alpha^{ext}(\vecr)=Y_{LM}(\hat{\vecr})v_{\alpha L}(r),\\
&&\delta\rho_\alpha(\vecr,\omega)=
  Y_{LM}(\hat{\vecr}){1\over r^2}\delta\rho_{\alpha L}(r,\omega),
\end{eqnarray}
and the linear response equation
\begin{equation}\label{respeq2}
\delta\rho_{\alpha L}(r,\omega)=\int_0 dr'
R_{0,L}^{\alpha\beta}(r,r',\omega)\left(
{\del v_\alpha\over \del\rho_\beta}(r'){1\over r'^2}
\delta\rho_{\alpha L}(r',\omega)+v_{\alpha L}(r',\omega)\right).
\end{equation}
This equation can be solved numerically with use of 
the mesh representation of the radial coordinate in
a way similar to Ref.\cite{Shlomo}.
The strength function for the response to the external field 
is expressed by means of the RPA response function as
\begin{eqnarray}
S(\hbar\omega)&&=\sum_k |\bra{0}\Vhat^{ext}\ket{k}|^2 
\delta(\hbar\omega-\hbar\omega_k)  \nonumber \\
&&=-{1\over \pi}{\rm Im}\int\int_0 dr dr' 
\sum_{\alpha\beta}
R_L^{\alpha\beta}(r,r',\omega)v^*_{\alpha L}(r)v_{\beta L}(r')
=-{1\over \pi}{\rm Im}\int_0 dr \sum_\alpha 
v^*_{\alpha L}(r)\delta\rho_{\alpha L}(r,\omega).
\end{eqnarray}

Let us discuss characteristic features of our linear
response theory by comparing with other approaches. Firstly,
the present formalism includes the continuum quasiparticle states
wherever they appear. Previous continuum QRPA's (or 
the continuum quasiparticle linear response theories)
\cite{Pl88FFSth,Bo96FFSGT,Kam98FFSe1,Hagino}
adopt approximations that take into account the continuum boundary condition
only for the normal Green function $G_0(E)$, but not for the abnormal 
Green function $F_0(E)$, while we have treated the 
single-quasiparticle HFB Green function exactly by using the 
construction by Belyaev et al.\cite{Belyaev}. 
We also emphasize that the present theory includes the configurations
where two nucleons are both in the continuum states.
Since the HFB single-quasiparticle spectrum is 
made of the discrete bound orbits and the continuum unbound
states, two-quasiparticle states are classified in three groups.
The first is the one where both quasiparticles occupy 
two discrete bound states. The second is where one quasiparticle
is in a  bound state with discrete excitation energy $E_i$ 
while the second quasiparticle occupy the continuum unbound states whose
excitation energy $E$ exceeds $|\lambda|$. 
These one-particle continuum states emerge above a 
threshold excitation energy $E_{th,1}=min E_{i}+|\lambda|$, where
$min E_i$ is the minimum quasiparticle excitation energy. 
In the unpaired limit, this threshold energy becomes
identical to the excitation
energy $-e_{h,last}$ of nucleon from the last occupied
hole orbit to the zero-energy threshold. 
The last group of two-quasiparticle states is the one where
both two quasiparticles occupy continuum unbound states. 
The threshold excitation
energy for the two-particle continuum states is 
$E_{th,2}=2|\lambda|$, i.e. it is twice the one-particle separation
energy $S_1$.
The present theory includes all the three groups as seen in Eq.
(\ref{uresp6}), whereas the previous
approaches include only the configurations where only one nucleon is
in the continuum states. 
Escaping processes of one particle to the external region
takes part in at the excitation energy $\hbar\omega >E_{th,1}$ above the
one-particle threshold, and consequently the strength functions show 
continuum spectra.  At excitation energy $\hbar\omega >E_{th,2}$ 
above the two-particle threshold, 
processes of two-particle escaping contribute.
A related point is that the previous continuum QRPA's adopt
the BCS approximation that has difficulty to describe 
the pairing in the continuum,
whereas we have exploited the coordinate space HFB formalism to 
remove the shortcoming.

Secondly, the present theory takes into account 
the particle-particle correlations associated with 
fluctuations in the pair densities
$\delta\rhot_+$, and $\delta\rhot_-$, which always emerge in 
paired systems. In other words, the pairing residual interaction
contributes to the linear response equation (\ref{respeq2}).
This particle-particle correlation has been neglected
in many continuum quasiparticle linear response theories or QRPA's. 
Even when it is included \cite{Pl88FFSth,Bo96FFSGT}, its effect 
is not clarified. We emphasize that this contribution 
should be included to keep a selfconsistency of 
the pairing correlation, and it indeed has important consequences.
This point is discussed in detail in the next section.

Thirdly, the present theory reduces to the continuum linear
response theory for unpaired systems \cite{Shlomo}
if the HFB ground state has the zero pairing potential.
This is easily seen in Eq.(\ref{uresp6}), which, 
in the zero pairing limit, has only the discrete part
consisting of the occupied hole orbits $\phi_{nlj}$. 
Note also that only the 11 component 
(the normal Green function $G_0$) of $\calGz$ contributes 
to the response function
for the density operator $\rhohat(\vecr)$.
In this sense, the
present theory is an extension of Ref.\cite{Shlomo}
to paired systems described by the coordinate space HFB.

\section{Numerical Analysis}

We apply the above formalism to spin independent quadrupole 
excitations in open-shell oxygen isotopes near the neutron 
drip-line, where both the pairing correlation and the continuum effects
may play important roles. In the following, we mainly discuss
results for the drip-line nucleus ${}^{24}$O.
Since our purpose in this
paper is to clarify basic characteristics of the theory, rather than to
make a precise prediction, we adopt a simple model Hamiltonian
which consists of a spherical
Woods-Saxon potential 
and the residual two-body interactions. 

As a pairing force
we assume the density-dependent delta force \cite{Esbensen,Te95HFBsd} 
\begin{equation}\label{ddpair}
v_{pair}(\vecr,\vecr')={1\over2}V_0(1-P_\sigma)
(1-\rho(r)/\rho_0)\delta(\vecr-\vecr'),
\end{equation}
where $P_\sigma$ is the spin exchange operator.
The interaction strength depends on the position through 
the total nucleon density 
$\rho(r)=\rho_{\rm n}(r)+\rho_{\rm p}(r)$. With
$\rho_0=0.16{\rm fm}^{-3}$ (the saturation density),
the pairing force is more effective at low-density nuclear
surface than in the interior region. 
The density-dependent pairing force is widely adopted in the HFB calculations
not only for unstable isotopes
\cite{DobHFB2,Db00HFBap,Te96HFBnrich1,Te97HFBnrich2,Ya00HFBdef,MeRiRMFB}
but also for rapidly rotating deformed nuclei \cite{Te95HFBsd,Duguet}.

To obtain the ground state, we solve the radial HFB equation 
with the box boundary condition $\phi_i(r=R_{max})=0$ 
according to the procedure of Ref.\cite{DobHFB1}, and determine
selfconsistently the nucleon
density $\rho_q(r) (q=n,p)$, the pair density $\rhot_q(r)$ and
the pairing potential $\Delta_q(r)=V_0
(1-\rho(r)/\rho_0)\rhot_q(r)$.
For the Woods-Saxon potential we adopt 
the standard parameter set \cite{Blomq,BMI}.
We have used the radial mesh size $\Delta r=0.2$ fm, and the
cut-off at $R_{max}=20$ fm. To evaluate 
the densities and the pairing potential, 
we include all the quasiparticle states 
whose excitation energy is below $E_{max}=50$ MeV and with the 
orbital angular momentum $l_{max}\leq 7\hbar$. 
The pairing force strength $V_0$ is
chosen $V_{0,n}=V_{0,p}=520 {\rm fm}^{-3}$MeV 
so that the average neutron pairing gap
$\left<\Delta_n\right>$  in ${}^{18,20}$O
approximately agrees with 
the overall systematics $\Delta \approx 
12/\sqrt{A} \approx 2.8$ MeV \cite{BMI}. Here we define the 
average gap by 
$\left<\Delta_n\right>=\int d\vecr\rhot(\vecr)\Delta_n(\vecr)/
\int d\vecr\rhot(\vecr)$, which corresponds to 
$\left<\Delta_n\right>_{uv}=
\sum_\mu u_\mu v_\mu\Delta_\mu / \sum_\mu u_\mu v_\mu$ 
($\mu$ denoting the canonical basis) adopted recently in
the literature\cite{Bender,Duguet,Ya00HFBdef}. 
The calculated average neutron gap is 
$\left<\Delta_n\right>=2.74, 3.13, 3.30$, and 3.39 MeV
for ${}^{18,20,22,24}$O, respectively. 
Note that the pairing gap increases as approaching the
neutron drip-line line.
For comparison,
we use also the density-independent pairing interaction (the volume-type
pairing), which is
given by dropping the density-dependent term in Eq.(\ref{ddpair}). 
In this case,
we use $V_0=240 {\rm fm}^{-3}$MeV, $\rho_0=\infty$ so that the
calculated average neutron gap 
$\left<\Delta_n\right>=2.70, 2.69, 2.26, 1.64$ MeV
(for ${}^{18,20,22,24}$O) gives approximately the same value in
${}^{18,20}$O. The neutron pairing gap for the 
volume-type pairing decreases as increasing the neutron number, especially 
at the drip-line nucleus $^{24}$O. This trend is opposite
to that for the density-dependent pairing, and consequently the calculated
pairing correlation differs significantly for ${}^{22,24}$O located 
near drip-lines
even though the force parameter is adjusted in more stable ${}^{18,20}$O.
Apparently, the density-dependent pairing force favors the pairing
correlation in the low-density surface region which develops in
nuclei near the drip-line. This is a characteristic feature
of the density-dependent pairing \cite{DobHFB1,DobHFB2,Db00HFBap}.

The quadrupole response of the system is described by means of the
linear response formalism presented in the previous section. As the residual
interaction acting in the particle-particle channel, we use the same
pairing interaction as that used for the static HFB calculation in order
to keep the selfconsistency, which turns out very important in the following.
For the particle-hole channel, we assume 
the Skyrme-type delta force 
\begin{equation}
v_{ph}(\vecr,\vecr')=
\left(t_0(1+x_0P_\sigma)+t_3(1+x_3P_\sigma)\rho(r)\right)
\delta(\vecr-\vecr'),
\end{equation}
where we adopt the same parameter as Ref. \cite{Shlomo};
$t_0=f \times (-1100)  {\rm fm}^{3}$MeV, 
$t_3=f \times 16000  {\rm fm}^{6}$MeV, $x_0=0.5$ and $x_3=1$. Here the 
renormalization factor $f$ is adjusted 
by imposing a selfconsistency\cite{Shlomo}
that the calculated static polarizability for the isoscalar dipole field
$D_{1M}^{IS}=\sum_n rY_{1M}+\sum_p rY_{1M}$ becomes zero for each nucleus.  
The linear response equations are solved in the following way.
The unperturbed response function is obtained by evaluating
Eq.(\ref{uresp5}) where the contour integral is performed numerically 
with an energy step $|\Delta E|=\eps/16$.
We use a small smoothing constant $\eps=0.2$ MeV, which
corresponds to a smoothing with a Lorentzian function with
FWHM of 0.4 MeV. As for 
the HFB Green function we adopt the exact expression (\ref{greenex}), and
evaluate the radial functions (also the response functions) by using 
the mesh $\Delta r=0.2$fm and the cut-off radius 
$R_{max}=20$fm.
In evaluating Eq.(\ref{uresp5}), we 
introduce also the angular momentum cut-off $l_{max}=7\hbar$.
These parameters are the same as in solving the static 
HFB calculation. 
The cut-off energy for the contour integral is $E_{cut}=50$ MeV, 
which is a natural choice since it is the same cut-off energy $E_{max}$
for the static HFB. Because the value of $E_{cut}$ is 
larger than the potential depth (measured from the Fermi energy), 
the contour integral can include all the hole orbits 
in the limiting case of the zero pairing potential. We thus 
guarantee that in the zero pairing limit 
the evaluated response function 
for the density operator becomes identical to the 
response function for unpaired systems \cite{Shlomo}.
The RPA response functions $R^{\alpha\beta}(r,r',\hbar\omega)$
is obtained by solving Eq.(\ref{respeq2}), 
which is a $600 \times 600 $ linear matrix equation
in the radial mesh representation.

\begin{figure}[tbp] 
\centerline{\psfig{figure=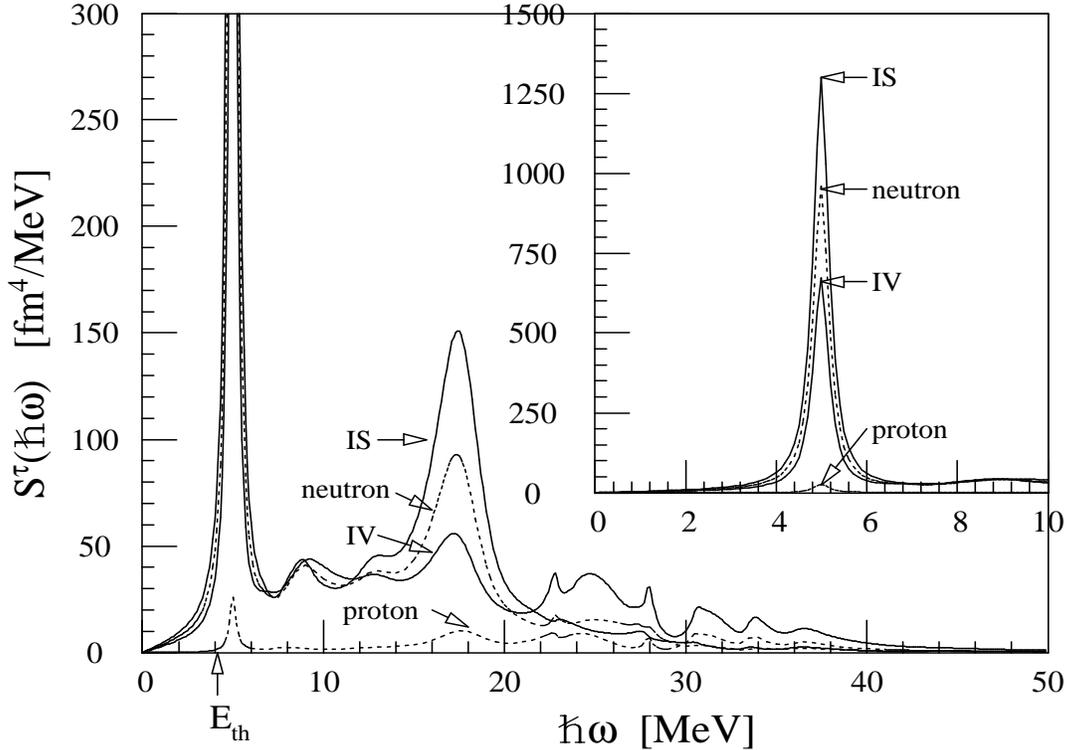,angle=-90,width=14cm}}
\caption{\label{Fig:str}
The strength functions for the quadrupole moments
$Q^{\tau}_{2M} \ (\tau=IS,IV,n,p)$
in $^{24}$O with use of the
density-dependent pairing force. 
The renormalization factor
for the particle-hole residual interaction is $f=0.632$.
Solid and dotted curves represent
those for $Q^{IS}_{2M},Q^{IV}_{2M}$, and $Q^{n}_{2M}, Q^{p}_{2M}$,
respectively. 
The inset shows a magnified part
for the low excitation energy $\hbar\omega<10$MeV.  The 
threshold energy $E_{th,1}=E_{th,2}=2|\lambda|=4.14$ MeV is indicated
by an arrow.
}
\end{figure}

Figure \ref{Fig:str} shows the strength functions calculated 
for the quadrupole excitation in $^{24}$O. Here we show the
strength function 
$S^{\tau}(\hbar\omega)=dB(Q^\tau 2, 0_{gs}^+\rightarrow k)/d(\hbar\omega)
=\sum_{Mk} \bra{0^+_{gs}}Q_{2M}^\tau\ket{k}^2
\delta(\hbar\omega-\hbar\omega_k)$
for the neutron and proton quadrupole moments
$Q^{n}_{2M}=\sum_n r^2Y_{2M}$ and 
$Q^{p}_{2M}=\sum_p r^2Y_{2M}$, the
isoscalar  and isovector quadrupole moments 
$Q^{IS}_{2M}=Q^{n}_{2M}+Q^{p}_{2M}$, and
$Q^{IV}_{2M}=Q^{n}_{2M}-Q^{p}_{2M}$
as a function of the excitation energy $\hbar\omega$.
Plotted here is the strength functions summed over the magnetic
quantum number $M$ so that their energy integral are
equivalent to $B(Q^\tau 2, 0^+_{gs}\rightarrow 2^+)$. 
A noticeable feature seen in Fig.\ref{Fig:str} is presence of
a sharp and intense peak around $\hbar\omega\approx 5.0$ MeV. 
There also exist 
a resonance peak around $\hbar\omega \approx 17$MeV with
a width of $\approx 3$MeV, and in addition a very broad distribution around
$\hbar\omega\approx 20-40$ MeV. 
These two modes correspond to isoscalar(IS) and
isovector(IV) giant quadrupole resonances (GQR) whereas they are not
pure ISGQR nor IVGQR since a large admixture 
of isovector (isoscalar) strength is present in each resonance
region of the isoscalar (isovector) modes.  

The peak excitation energy of the low-lying mode is $\hbar\omega=5.0$MeV.
Since the threshold energies 
for one-neutron and two-neutron continuum states are, 
$E_{th,1}=E_{th,2}=4.14$MeV, the low-lying neutron quadruple mode
is embedded in the neutron continuum states.
The FWHM width of this mode $\approx 45$keV is almost the same as the
smoothing width $2\eps=40$keV,
indicating that the mode has very small escaping width in spite of 
the coupling to the neutron continuum states.  

The low-lying quadrupole mode at $\hbar\omega\approx 5.0$MeV is characterized
by very large neutron strength. The isoscalar strength is also large, but
the proton strength is small. The isovector strength is sizable but
smaller than the neutron and isoscalar strengths because the small
proton contribution is in phase with the neutrons.
The integrated neutron strength $B(Q^n 2)$ below $\hbar\omega=6.0$ MeV is 
$605 {\rm fm}^{4}$ whereas $B(E2)(=e^2B(Q^p 2))$ is
only $16.4 e^2{\rm fm}^{4}$.
Converting the strengths to the neutron proton ratio 
$M_n/M_p$ for the quadrupole
transition amplitudes, it is evaluated as $M_n/M_p= 6.1$.
The neutron character of this mode is apparent by comparing this value to
a simple macroscopic estimate $M_n/M_p=N/Z=2$.
The low-lying mode exhausts approximately $20\%$, $10\%$  and 
$19\%$, of the sum rule value for the isoscalar,
isovector, and neutron  quadrupole transition strength, respectively
(See Eq.(\ref{sumrule}) and Fig. \ref{Fig:sum-rule}(a)).
This ratio is
much larger than a typical value for the low-lying isoscalar quadrupole mode 
($\sim 10\%$) \cite{BMII}. On the contrary, the 
proton strength (i.e. the $B(E2)$ strength) is small, 
exhausting only $2\%$ of the energy weighted sum rule. 
The strength of the low-lying neutron mode is continuated to the
neutron strength distribution in the interval 
($\hbar\omega\approx 6-15$ MeV) between the low-lying
mode and the giant resonances, which corresponds to 
the so called threshold strength. However, the character of the
low-lying neutron mode clearly differs from that of the threshold
strength \cite{HaSaZh} since its large neutron strength
of collective nature is generated by the residual interaction (see
below).  We note also that 
the HFB single-quasiparticle states have no discrete bound orbits, instead they
are all continuum states or resonances. The low-lying neutron mode is 
a collective state made of continuum two-quasiparticle states.

We have also calculated the quadruple strength functions for
$^{18,20,22}$O. The peak energies of the low-lying mode is 
$\hbar\omega=4.2, 4.1, 4.5, 5.0$ MeV for $A=18,20,22,24$, respectively, 
which is $1-2$ MeV higher than the experiments
\cite{Khan-etal,BE2,Jewell,Thirolf,Azaiez}. 
The calculated neutron
strength ($B(Q^n 2)=152,295,426,605 {\rm fm}^{4}$ for A=18-24) 
increases sharply with increasing the neutron number.
On the contrary, the calculated $B(E2)=15,18,17,17 e^2{\rm fm}^{4} (A=18-24)$
stays almost constant. The calculation
underestimates the experimental $B(E2)$ value 
\cite{BE2,Jewell,Khan-etal,Thirolf}
by a factor of 3-1.5 for
$^{18,20}$O, but with reasonable agreement in $^{22}$O.
The calculated $B(E2)$ values differs by a factor of 0.5-1.5 from the 
quasiparticle RPA calculation in Ref.\cite{Khan-Giai,Khan-etal}. 
The adopted model
Hamiltonian, especially the Woods-Saxon potential,
should be improved to 
make more quantitative comparison, e.g. by using more realistic
Skyrme interaction both for the Hartree-Fock potential and 
for the particle-hole interactions. 
Leaving such improvements for a future investigation,
we in the following focus on basic aspects of
the present theory.

\begin{figure}[tbp]  
\centerline{\psfig{figure=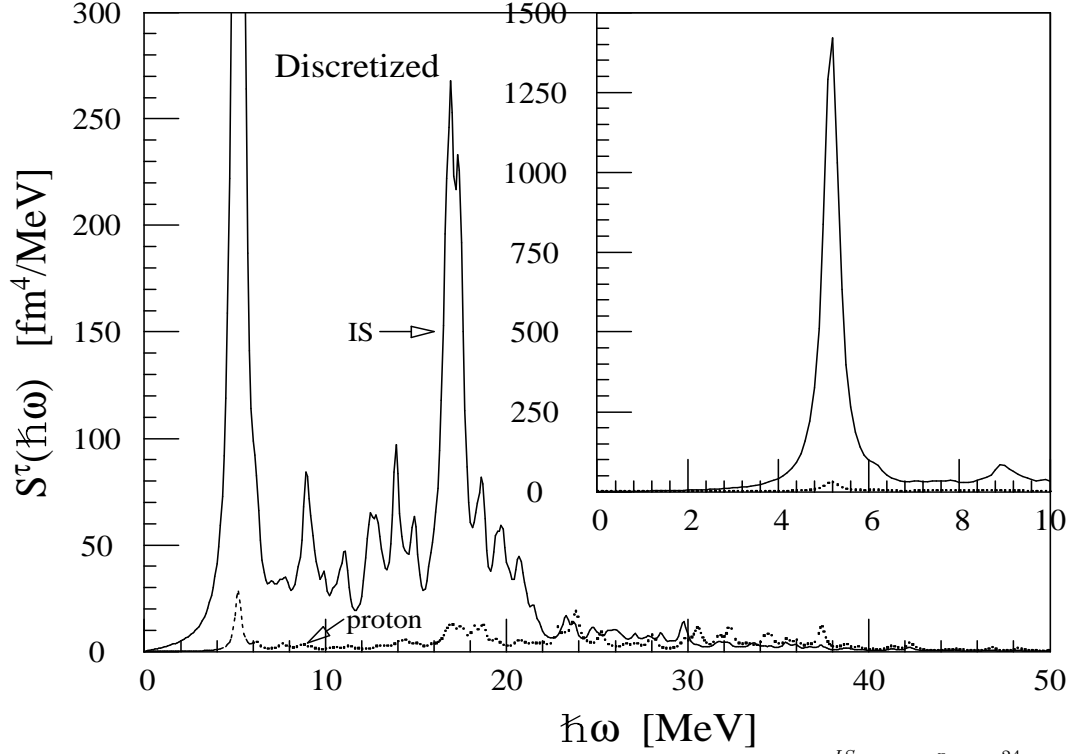,angle=-90,width=14cm}}
\caption{\label{Fig:discrete}
The strength functions for the isoscalar and proton
quadrupole moments $Q^{IS}_{2M}$ and $Q^{p}_{2M}$
in $^{24}$O with use of the density-dependent pairing force,
calculated with the discretized continuum states (see text). 
The renormalization factor
for the particle-hole residual interaction is $f=0.702$.
Solid and dotted curves represent
those for $Q^{IS}_{2M}$ and $Q^{p}_{2M}$, respectively. 
}
\end{figure}

The present theory takes into account the one- and two-particle continuum 
states without making any discretization method or bound state approximation.
It is interesting in this respect to see difference between the
results presented above and 
results that would have been obtained by using discretized 
continuum states. We show in Fig.\ref{Fig:discrete} the strength 
function obtained by using the discrete spectral representation 
Eq.(\ref{uresp2}) instead of Eq.(\ref{uresp5}).  
Here we solve discrete single-quasiparticle states 
with the box boundary condition $\phi_i(R_{max})=0$, and 
adopt all the quasiparticle states 
with the excitation energy $E_i$ below 
the energy cut-off $E_{max}=50$ MeV and the angular momentum
cut-off $l_{max}=7$, including the discretized states
in the continuum region $E_i>|\lambda|$. The renormalization
factor $f$ is slightly changed  in this approximate calculation 
in order to reproduce the zero energy dipole mode.
With use of the discretized continuum approximation, 
the strength function consists only of discrete states (but smeared with
the smoothing width) even in the continuum region above the
threshold. This causes
spurious fluctuation, as seen in Fig.\ref{Fig:discrete}. 
It is noted on the other hand that
a gross profile of the 
strength distribution is described fairly well by the discretizing
approximation. It can be expected that in the limit of large
box size $R_{max} \rightarrow \infty$, the discretizing approximation would
give the same results with the one obtained with the continuum
linear response theory, whereas such limiting is practically very 
difficult. Note also that the strengths associated with the low-lying neutron 
mode at $\hbar\omega\approx 5$ MeV are almost reproduced by the discretizing
approximation. This is because the low-lying mode in the present 
model calculation  is a very narrow resonance, for which
a bound state approximation can be a reasonable approximation.

\begin{figure}[tbp]  
\centerline{\psfig{figure=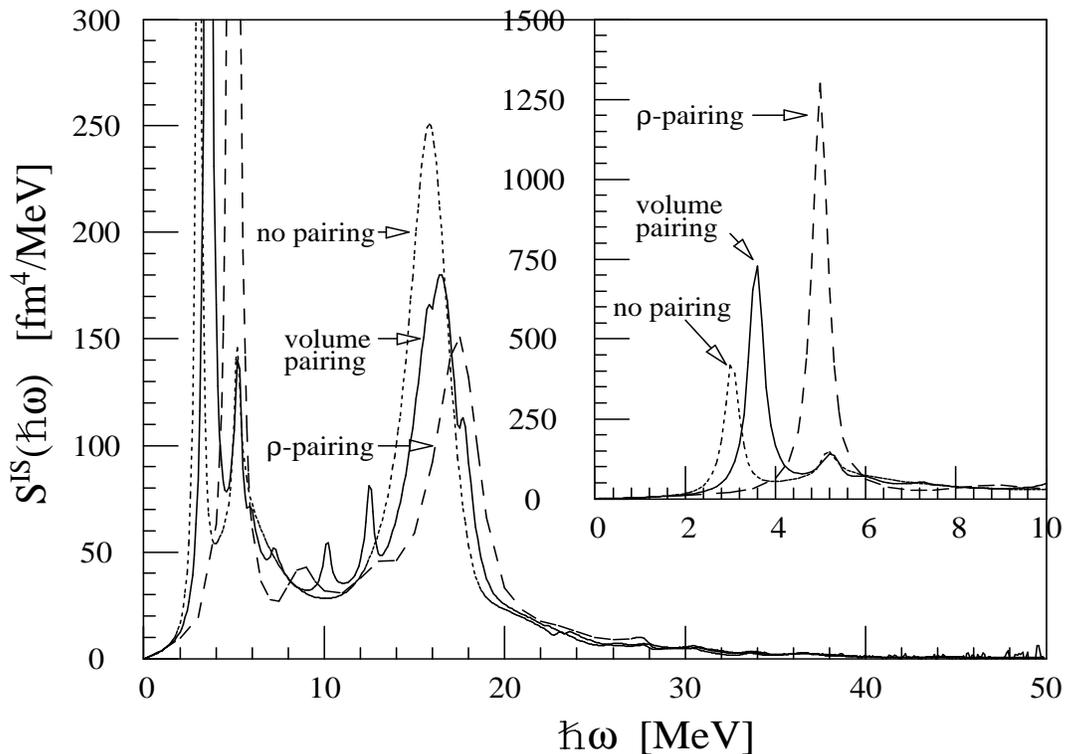,angle=-90,width=14cm}}
\caption{\label{Fig:pair-dependence}
Dependence of the quadrupole strength function in $^{24}$O 
on the pairing interaction.  Plotted are
the strength functions for the isoscalar
quadrupole moment $Q^{IS}_{2M}$ for the density-dependent pairing
(dashed curve), for the volume-type pairing (solid), and
for the case where the pairing is neglected (dotted). See text for details.
The renormalization factor
for the particle-hole residual interaction is $f=0.702$.
}
\end{figure}

The pairing correlation play various important roles for 
the response. We show in
Fig.\ref{Fig:pair-dependence} 
the calculation without the pairing correlation (i.e. the pairing
interaction is neglected both in the static HFB and the linear
response equation). 
In this case,  the quadrupole strength in the low-energy region
becomes much smaller. The small peak at $\hbar\approx 3.0$ MeV 
is not very collective, but it is basically non-collective
neutron particle-hole transition  $2s{1\over 2}\rightarrow 1d{3\over2}$
with slight enhancement due to correlation.
It is clearly seen that
the pairing increases the collectivity of the low-lying neutron mode. 
The pairing effect on the low-lying isoscalar quadrupole mode 
in stable nuclei is well known\cite{BMII}. Our result suggests
a similar effect on the neutron mode, which
is embedded in the continuum states in the case of drip-line nuclei.

We found further that the low-lying neutron mode is
quite sensitive to the density-dependence of the pairing. 
Figure \ref{Fig:pair-dependence} shows also the result
that is obtained with the density-independent volume-type 
pairing force  in
place of the density-dependent one Eq.(\ref{ddpair}).
The neutron (and isoscalar) strength in
the low-lying neutron mode obtained in this calculation  is 
significantly small, i.e. it is about 50\%
of that with the density-dependent pairing. 
Using this sensitivity, one may be able to probe the
density-dependence of the pairing.
The difference between the density-dependent and the volume 
pairing is most significant in $^{24}$O, but not very large in
the other oxygen isotopes,
which is in accordance with the behavior of the average neutron
gap $\left<\Delta_n\right>$ discussed above. 
Fig. \ref{Fig:pair-dependence} shows
that the pairing correlation influences also 
the strength distribution in the giant resonance
region. It has an effect to reduce the strength in the 
high lying modes.
An apparent reason is that
the low-lying neutron mode collects nearly $20\%$ 
of the total energy weighted isoscalar and neutron sums 
in the case of the surface pairing, and
removes corresponding amount of strength from the giant resonance region
because of the sum rule.
The peak position and width
(distribution profile) of GR's are also slightly affected by the
pairing.

\begin{figure}[tbp]  
\centerline{\psfig{figure=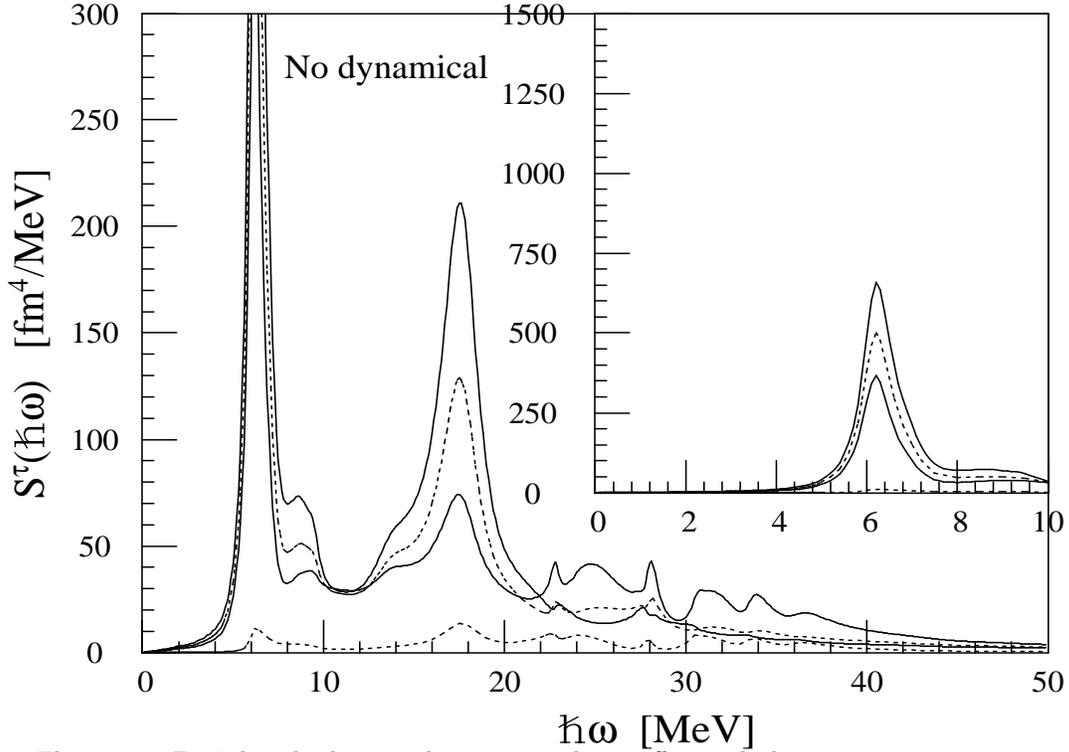,angle=-90,width=14cm}}
\caption{\label{Fig:pair-partial}
The same as Fig.\ref{Fig:str}, but the dynamical pairing
correlation effect in the linear response equation 
is neglected. Namely the density-dependent pairing 
is taken into account only in the static HFB mean-field.
}
\end{figure}

There are two kinds of mechanisms in the pairing effects on 
the collective excitations. 
The first is a static effect. Since 
the static pair potential $\Delta(r)$ in the 
HFB mean-field influences strongly the quasiparticle excitations,
it hence affects the collective excitations. The second is dynamical.
It is a correlation effect that is caused by the residual pairing
interaction entering in the linear response equation (\ref{respeq2}).
This arises because the collective excitations in the
paired system induce not only 
the fluctuation in the density $\delta\rho$ but also 
the fluctuations in the pair densities $\delta\rhot_+$ and $\delta\rhot_-$,
for which correlation is brought by the residual
pairing interaction (see, Eq.(\ref{respeq2})). 
To quantify this dynamical pairing correlation effect,
we show in Fig.\ref{Fig:pair-partial} 
a calculation in which the pairing interaction is
neglected in the linear response equation (\ref{respeq2}),
but included in the static HFB mean-field. 
Comparison with Fig.\ref{Fig:str} shows that 
the dynamical pairing correlation has a sizable effect to lower 
the excitation energy of 
the low-lying neutron mode by about 1MeV. 
(One may also note that the width of the low lying mode
increases slightly in this calculation. ) Performing a calculation
where the particle-hole interaction is neglected,
we find that the low-lying neutron mode is produced also by 
the pairing interaction alone, although the particle-hole 
residual interaction together is essential to make the collectivity large. 
If we used the 
the conventional
monopole pairing force (the seniority force) instead of the
zero-range pairing force, the 
dynamical pairing correlation effect would be missed.
The dynamical effect is rather related to effects of the
quadrupole pairing that has been discussed in connection
with the low-lying quadrupole mode in stable nuclei.

\begin{figure}[tbp]  
\centerline{\psfig{figure=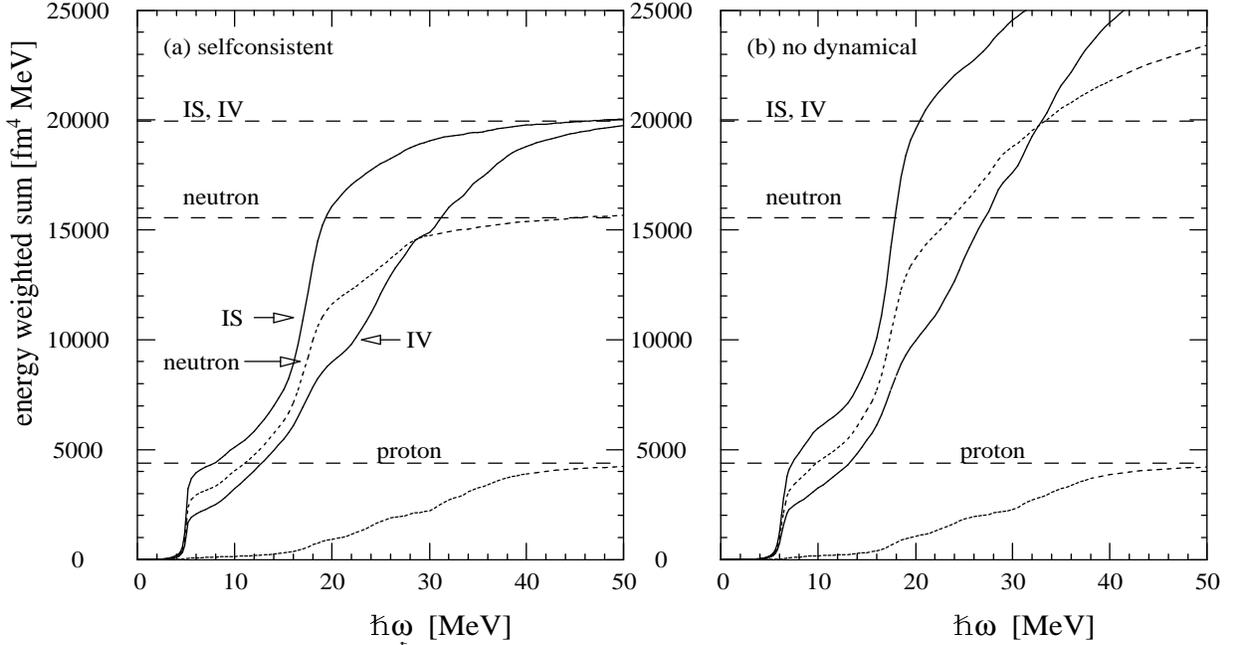,angle=-90,width=16cm}}
\caption{\label{Fig:sum-rule}
(a)Energy weighted sum 
$\int_0^{\hbar\omega} \hbar\omega' S^{IS,IV,n,p}(\hbar\omega')\hbar d\omega'$ 
of the quadrupole strength functions in $^{24}$O, shown in Fig.\ref{Fig:str}. 
Here the density-dependent pairing is taken into account 
selfconsistently both in the
static HFB mean-field and in the linear response equation.
The dashed horizontal lines show the sum rule values for 
isoscalar (the same for isovector) transitions, and those 
for the neutron and the proton transition strengths.
(b) The same as (a), but the dynamical pairing effect is
neglected (cf. Fig.\ref{Fig:pair-partial}). 
Namely the pairing is taken into account 
only in the static HFB mean-field.
}
\end{figure}

One of the most important characteristics of the present theory is that 
the energy weighted sum rule 
\begin{mathletters}\label{sumrule}
\begin{eqnarray}
&\int_0^{\infty} \hbar\omega S^{IS,IV}(\hbar\omega) \hbar d\omega
= {25\hbar^2\over 4\pi m}
\left( N\left<r^2\right>_n + Z\left<r^2\right>_p \right), \\
&\int_0^{\infty} \hbar\omega S^{n,p}(\hbar\omega) \hbar d\omega
= {25\hbar^2\over 4\pi m}
N\left<r^2\right>_n, \ \  {25\hbar^2\over 4\pi m}Z\left<r^2\right>_p,
\end{eqnarray}
\end{mathletters}
is satisfied very accurately. 
The sum rule should hold since the adopted residual interactions (both
the particle-particle pairing force $v_{pair}$ and the particle-hole
interaction $v_{ph}$) keep the Galilei invariance \cite{BMII}, i.e.,
they commute with the
quadrupole operator $r^2Y_{2M}(\hat{\vecr})$. 
This is demonstrated in Fig. \ref{Fig:sum-rule}(a) that
shows the calculated energy weighted sum $\int_0^{\hbar\omega} 
\hbar\omega' S^\tau(\hbar\omega') \hbar d\omega' \ \ (\tau=IS,IV,n,p)$
as a function of the energy boundary $\hbar\omega$.
We stress here that the selfconsistent treatment 
of the pairing both in the static HFB  and in the dynamical linear
response is crucial to satisfy the sum rule.
This is shown in Fig. \ref{Fig:sum-rule}(b). This plots
the energy weighted sum obtained in a truncated 
calculation in which the residual pairing interaction is neglected
in the linear response equations (cf. the corresponding strength function
is shown in Fig.\ref{Fig:pair-partial}). 
In this calculation, the sum rule Eq.(\ref{sumrule}) is
strongly violated by about 50\% for neutrons. 
The violation is significant especially in the high lying region
($E \gesim 10$ MeV), as the strength in this region
(shown in Fig.\ref{Fig:pair-partial}) apparently overestimates
that of the full calculation (Fig.\ref{Fig:str}).
The violation originates from the inconsistent treatment
of the pairing interaction in the calculation of
Figs.\ref{Fig:pair-partial}
and \ref{Fig:sum-rule}(b), where
the pairing correlation
is included only in the HFB static mean-field, but neglected in
the linear response equation.  One can understand this by noting
that the static pair potential $\Delta(r)$ violates the 
Galilei invariance,i.e., it does not commute with the density operator. 
By taking into account selfconsistently the 
pairing interaction in the linear response equation, 
the Galilei invariance of the original pairing interaction is
recovered, and the sum-rule is satisfied. 
This corresponds to a previous result\cite{Kubo} showing that
the sum rule violated by the monopole pairing potential can
be remedied by including the selfconsistent quadrupole pairing interaction
that recovers the Galilei invariance.
The selfconsistent
treatment of the pairing is important especially in the case
of the density-dependent pairing force and in nuclei near drip-lines
since in such cases the pairing potential $\Delta(r)$ 
in the surface region becomes relatively large.

We have also checked that the selfconsistency in the pairing is fulfilled
in the present calculation, by investigating the Nambu-Goldstone mode in
the response functions for the monopole pair operators $P_0^\dagger=\int
d \vecr P^\dagger(\vecr)$ and
$P_0=(P_0^\dagger)^\dagger$. 
When the HFB static pairing potential $\Delta(r)$ is
not zero, the pairing rotational mode that has  zero excitation
energy should emerge as a Nambu-Goldstone mode associated with
the nucleon number conservation of the Hamiltonian, or in other word  
the invariance with respect to the gauge rotation 
$e^{i\theta \hat{N}}$ \cite{Ring-Schuck}
($\hat{N}=\int d\vecr\sum_\sigma \psid(\vecrs)\psi(\vecrs)$ being 
the neutron or proton number operator ). The
monopole response calculated for $^{24}$O 
exhibits the pairing rotation mode at the excitation energy 
very close to zero. It is found that 
we can move the energy of the pairing rotational mode exactly at zero energy
just by modifying tinily the pairing force strength $V_0$ within $1\%$.
We also checked that the calculated 
monopole strength function for the nucleon number operators $\hat{N}$ 
carries no spurious strengths. This is again achieved by including
selfconsistently the dynamical pairing correlation 
in the linear response equation.

\section{Conclusions}

We have formulated a new continuum linear response theory on the
basis of the Hartree-Fock-Bogoliubov formalism in the
coordinate space representation. 
This enables us to describe the pairing correlation in nuclei
near drip-lines in a selfconsistent way both 
in the static ground state and in the dynamical collective responses.
The formalism is able to 
include effects of the one-particle and the two-particle continuum
states on the collective excitations. 

We have described the quadrupole response in 
the drip-line nucleus $^{24}$O with use of the density-dependent 
zero-range interactions. The strength functions for the quadrupole
transition moments are obtained up to the giant resonance region.
A low-lying mode which has significant neutron collectivity and 
is embedded in the neutron continuum states is obtained. 
We have analyzed in detail pairing effects important
for the low-lying neutron mode. It is found that
the low-lying neutron mode is sensitive to the density-dependence 
of the pairing correlation especially near the drip-line. 
The collective excitations in paired systems induce 
fluctuations not only in the normal density but also in the pair
densities. The residual part of the pairing interaction 
causes dynamical correlation effects on the responses through the pair density
fluctuations. The present theory describes selfconsistently 
both the static pairing effect caused by 
the HFB pair potential and the dynamical pairing correlation
in the linear responses. The energy weighted sum rule is satisfied
very accurately. This is because the selfconsistent treatment
of the pairing restores the Galilei invariance,
which would be violated if the HFB pair potential alone was included.

\section*{Acknowledgments}

The author thanks K. Matsuyanagi for 
valuable comments and discussions.

\end{document}